# Accurate measurement of the cutoff wavelength in a microstructured optical fiber by means of an azimutal filtering technique

**Laurent LABONTE, Dominique PAGNOUX, Philippe ROY**

*Xlim/ Photonic department, UMR CNRS 6172 and University of Limoges*

*123Av. A Thomas 87060 Limoges, France*

**Faouzi BALHOUL, Mourad ZGHAL**

*ENIT, BP37, Le Belvédère, Tunis, Tunisia and Sup'Com, 2083 El Ghazala Ariana, Tunisia*

**Gilles MELIN, Ekaterina BUROV**

*ALCATEL Research and Innovation, Route de Nozay, 91460 Marcoussis , France*

**Gilles RENVERSEZ**

*Institut Fresnel, UMR CNRS 6133 and University Paul Cézanne Aix-Marseille III*

*Faculté des sciences de St Jérôme, 13397 Marseille, France*

## ABSTRACT

A simple self-referenced non destructive method is proposed for measuring the cutoff wavelength of microstructured optical fibers (MOFs). It is based on the analysis of the time dependent optical power transmitted through a bow-tie slit rotating in the far-field pattern of the fiber under test. As a first demonstration, the cutoff wavelength of a 2m MOF sample is measured with a precision of ±10nm, in good agreement with theoretical predictions.

Since their invention ten years ago microstructured optical fibers (MOFs) have attracted a lot of interest because they offer the possibility of managing the propagation characteristics of the guided light in a much larger range than in conventional fibers[1]. In particular, it has been demonstrated that MOFs can be endlessly single mode when the condition d/Λ~<0.42 is fulfilled, d being the diameter of the holes and Λ being the pitch of the hole lattice[2]. On the other hand, efficient supercontinuum generation and other non linear process have been achieved in MOFs, with a small core surrounded by one or few rings of large holes, in which the zero dispersion wavelength is shifted towards short wavelengths[3]. In such applications, the endlessly single mode condition is not fulfilled any more. This means that there is a cutoff wavelength, commonly called $\lambda_c$, under which higher order modes can propagate and which value must be precisely determined. The modal cutoff of MOFs has already been theoretically investigated by considering perfect but finite size structures[2]. These works point out the high sensitivity of $\lambda_c$ to the geometrical parameters of the fibers. Because actual manufactured fibers may suffer from unavoidable geometrical imperfections, it is necessary to achieve a reliable measurement of $\lambda_c$. The technique prescribed in the normalized procedure for standard fibers is based on the measurement of higher order modes differential loss versus fiber bending radius[4]. It is no more valid for characterizing MOFs because, due to the high NA of the fiber, higher order modes are much less sensitive to bending than the corresponding modes in standard fibers. An alternative method based on the measurement of the power transmitted over a long and a short length of fiber successively (cutback method) has been recently proposed[5]. The curve of the differential transmission versus the wavelength exhibits local maxima close to the cutoff of higher order modes, due to the large increase of their confinement loss. However, the difference between the two measured spectra remains weak, and this technique may suffer from very little changes in the launching conditions and from the source power or detector sensitivity drifts. In this letter, we report that is to our knowledge the first self referenced non destructive measurement of $\lambda_c$ into MOFs. The method is based on the azimuthal analysis of the far field pattern at the output of the tested fiber[6]. The experimental set-up is shown on the figure 1.

A dual wavelengths (1064/532nm) microchip pulsed Nd:YAG laser (repetition rate: 5.4kHz, pulse duration :0.6ns) pumping a specially designed 4m long MOF with a 800nm zero dispersion provides a bright continuum [300-1700nm] at the output, that is used as the source for our measurement[7]. The light from this fiber source (mean power=5mW) is coupled by a slightly misaligned butt joint connection into the core of a 2m sample of the fiber under test (FUT), providing the excitation of both the fundamental and the higher order modes. The far-field pattern from the output of the FUT is scanned by means of a rotating "bow-tie" slit, which is rotated at the frequency $f_s$=1.5Hz. The part of the light crossing the slit is focused by a concave mirror on the input face of a large core diameter multimode fiber. At the output of this collecting fiber, the signal is chopped at a frequency $f_c$=60Hz and it is spectrally filtered by a 0.5nm spectral width monochromator. Finally, the optical power at the output of the monochromator is detected by a InGaAs large sensitive area detector and the spectrum of the electric signal is displayed by a FFT spectrum analyzer.

The azimuthal symmetry of each mode diffracted at the output is preserved in the far-field. Thus, in the single mode regime, due to the $\pi/3$ symmetry of the fundamental mode, lateral lines at $f_c \pm 6f_s$ and possible harmonics at $f_c \pm n.6f_s$ ($n \in N$) can be observed on the displayed spectrum. In the bimode regime, the symmetry of $\pi$ of the first higher order modes in fabricated MOFs generates new lateral lines at $f_l = f_c \pm 2f_s$, which amplitude depends on the part F of the output power carried by this mode ($0 \leq F \leq 1$). In the normalized measurement method for conventional fibers (ITU-T G650), the criterion for determining $\lambda_c$ corresponds to F=0.0225 at the output of the FUT, assuming that the fundamental and the second modes are equally excited at the input[8]. In the case of conventional fibers, it has already been demonstrated in the ref.8 that $R(\lambda)=D(\lambda)$, in which :

$$R(\lambda)=10\log\frac{P_{11}}{P_{01}}=10\log\frac{F}{1-F} \quad \text{and} \quad D(\lambda)=10\log\frac{2A_l}{A_c-2A_l} \quad (1)$$

$P_{11}$ and $P_{01}$ being the power carried by the $LP_{01}$ mode and the $LP_{11}$ mode respectively and $A_c$ and $A_l$ being the linear amplitudes of the lines at $f_c$ and $f_l$ respectively. In these conditions, D may theoretically vary from $-\infty$ to $+\infty$ when F is increased from 0 (single mode regime) to 1 (propagation of the pure second mode in the bimode regime) and $D(\lambda_c)=-16.4$dB for F=0.0225. In the case of MOFs, because of the somewhat different spatial distribution of the power into the modes, $D(\lambda)$ becomes different from $R(\lambda)$, i.e. $D(\lambda_c) \neq -16.4$dB. A large set of simulations that we have carried out show that $D(\lambda_c)$ can vary in the range from approximately $-22$dB to $-14$dB according to the opto-geometrical parameters of the considered MOF (to be published further). Nevertheless, when crossing the cutoff wavelength, F increases from 0 to small but significant values and D suffers a very abrupt variation, allowing a precise determination of $\lambda_c$. The method has been applied to a MOF manufactured by Alcatel using the standard stack and draw process. A SEM picture representative of the characterised 2m sample is given in the inset of Fig. 1. Its average geometrical parameters, respectively $\Lambda=2.60$ µm and d=1.43 µm (d/$\Lambda$=0.55) have been evaluated thanks to a commercial image analysis system. The figures 2a and 2b show two typical spectra displayed by the spectrum analyzer respectively in the single mode domain (D=-20dB) and in the bimode regime (D=-8dB). The undesirable lines at $f_c \pm f_s$ are due to the slight residual unbalance in the rotation of the spatial filter. The effect of this alignment imperfection is obviously more significant (higher lines) in the presence of the second mode because of the rapid radial variation of its intensity near the center. The curve $D(\lambda)$, deduced from the lines at $f_c$ and at $f_l$ in the measured spectra, is reported on the figure 3. It exhibits a 10dB abrupt step when $\lambda$ is decreased from 1380nm to 1360nm, allowing to precisely locate $\lambda_c$ at 1370nm±10nm. This result has been compared to the value of $\lambda_c$ that has been theoretically determined for an ideal finite size MOF ($\Lambda$=2.6µm and d=1.43µm) mimicking the fabricated one, as the wavelength $\lambda_{c\text{-th}}$ for which the parameter $Q(\lambda) = \dfrac{d^2 \log[\text{Im}(n_{\text{eff}})]}{d^2 \log(\lambda)}$ exhibits a sharp minimum, $n_{\text{eff}}$ being the complex effective index of the considered higher order mode[2]. Two methods have been used to compute $n_{\text{eff}}(\lambda)$ : a classical vectorial finite element method (FEM) with perfect matched layers (PMLs) at the outer boundaries of the modelized structure[9], and the well-established multipole method (MM)[10]. On the curves of $Q(\lambda)$ reported in Fig. 4, we can notice a ~15nm difference between the $\lambda_{c\text{-th}}$ obtained by the two methods for each of the four EM higher order modes ($TE_{01}$ like, $TM_{01}$ like, a degenerate pair $HE_{21(x,y)}$ like). This discrepancy is attributed to the influence of the PML in the FEM, no such technique being needed with the MM. However the two methods provide similar results. $\lambda_{c\text{-th}}$ of the $TM_{01}$ like mode is found about 85 nm and 70nm higher than that of $HE_{21}$ like and $TE_{01}$ like modes, respectively. The theoretical cutoff region for these higher order modes extends from 1.39µm to 1.49µm. This result is close but slightly higher than the experimental one, as also noticed in the ref. 5. This is not surprising if we consider the two following points. First, the simulations only address the confinement loss of the ideal MOF and don't take into account the effects of the geometrical imperfections of the actual fiber. Second, the criterion used for the experimental determination of $\lambda_c$ is different from the one used in the simulations, and the two can't be directly related. At last, it is worth noting that we don't control the polarization of the light launched into the FUT. This means that we don't control the part of the propagating power that is carried by each higher order mode. The fact that their cutoff wavelengths are distributed over about 100nm could explain the irregular shape and the slow decrease of the curve $D(\lambda)$ between 1390nm and 1500nm.

In conclusion, we have proposed a new efficient self-referenced method for accurately measuring the cutoff wavelength of MOFs. The first reported measurements are in fairly good agreement with the theoretical predictions. Furthemore, this method makes it possible to study the length and the curvature dependence of $\lambda_c$ in MOFs (work in progress).

Acknowledgements : We are grateful to our colleagues Vincent Couderc et al. from Xlim for providing us a prototype of the continuum source that they have developed.

**Figure 1:** experimental setup (cross section of the microstructured fiber shown in the inset)

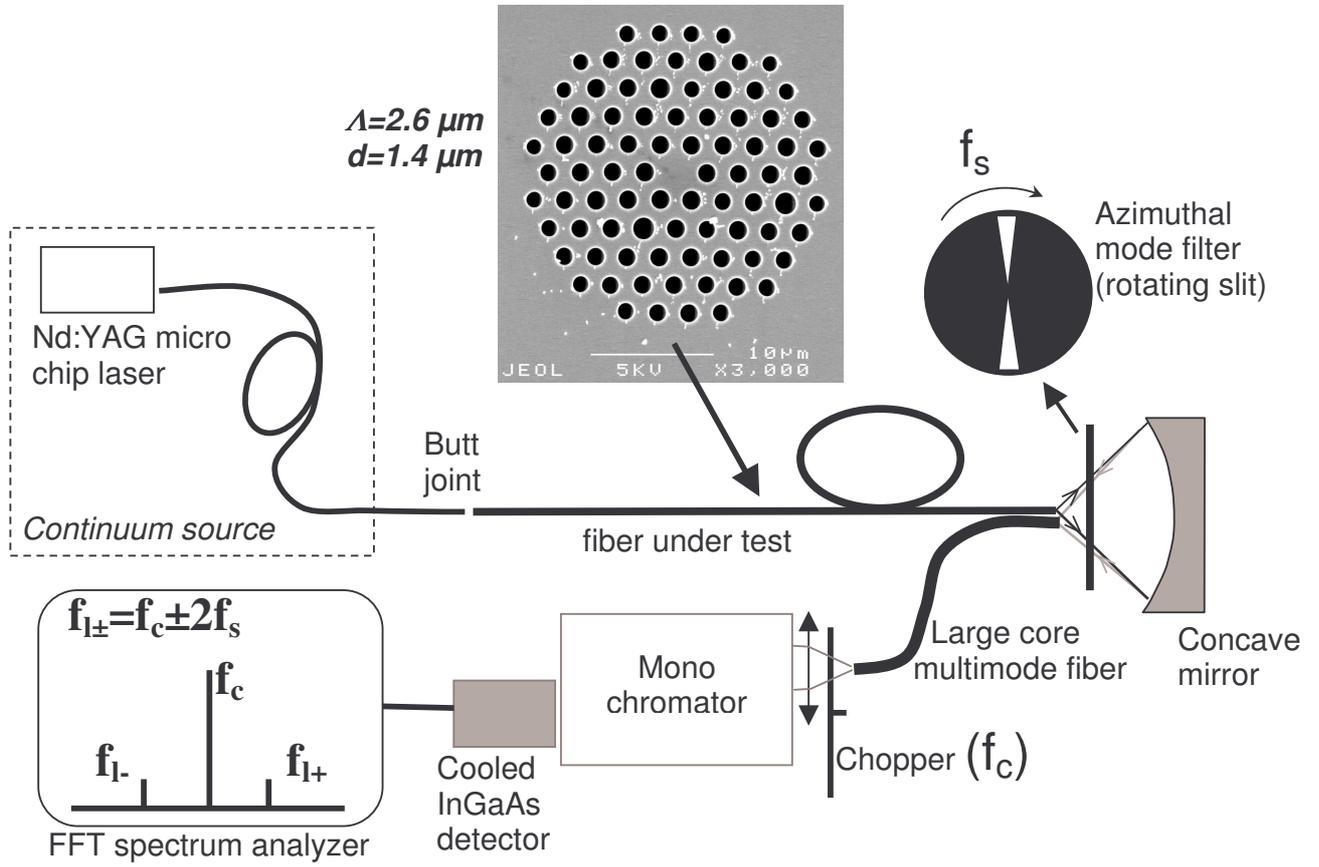

Figure 1

**Figure 2 :** typical spectra measured with the setup described in fig.1 a) single mode regime  b) bimode regime ($\Delta = A_c(dBV) - A_l(dBV)$))

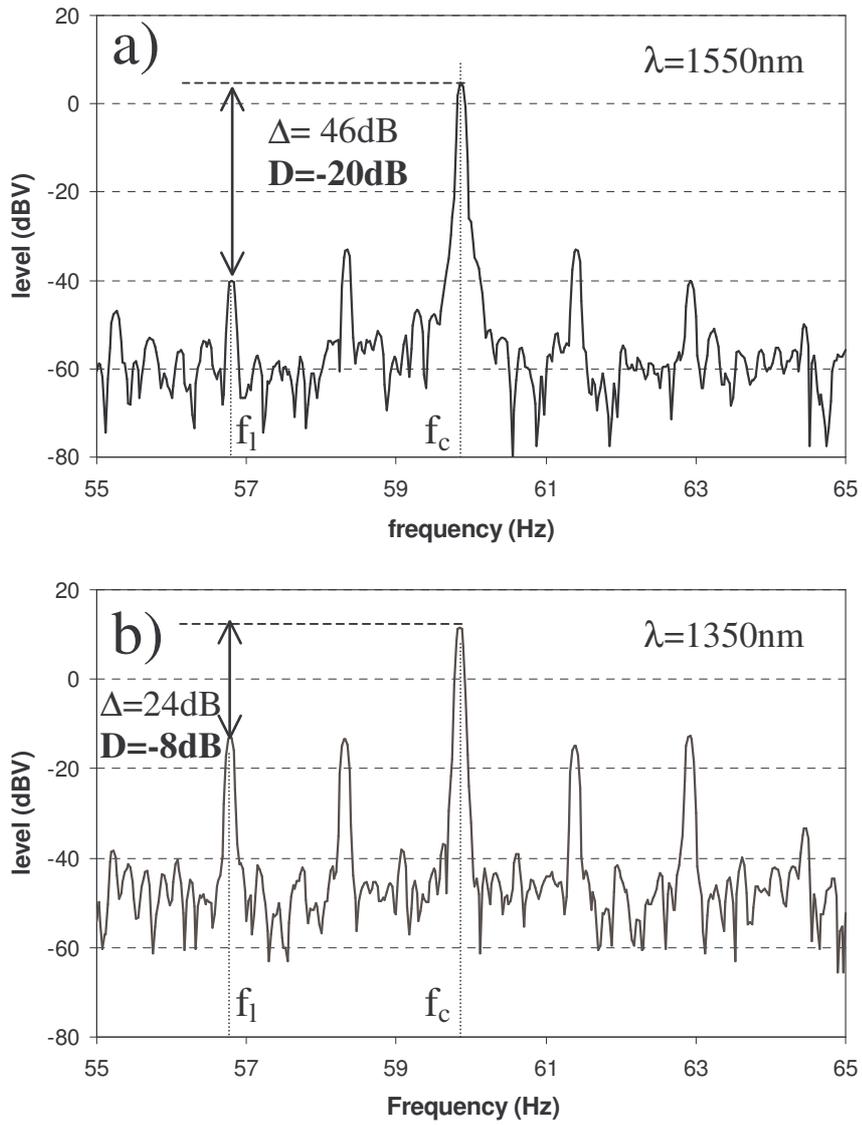

Figure 2

**Figure 3 :** Curve D(λ) measured with the fiber shown in the fig.1 and described in the text ($\lambda_c$=1370nm, $\Delta\lambda$=20nm)

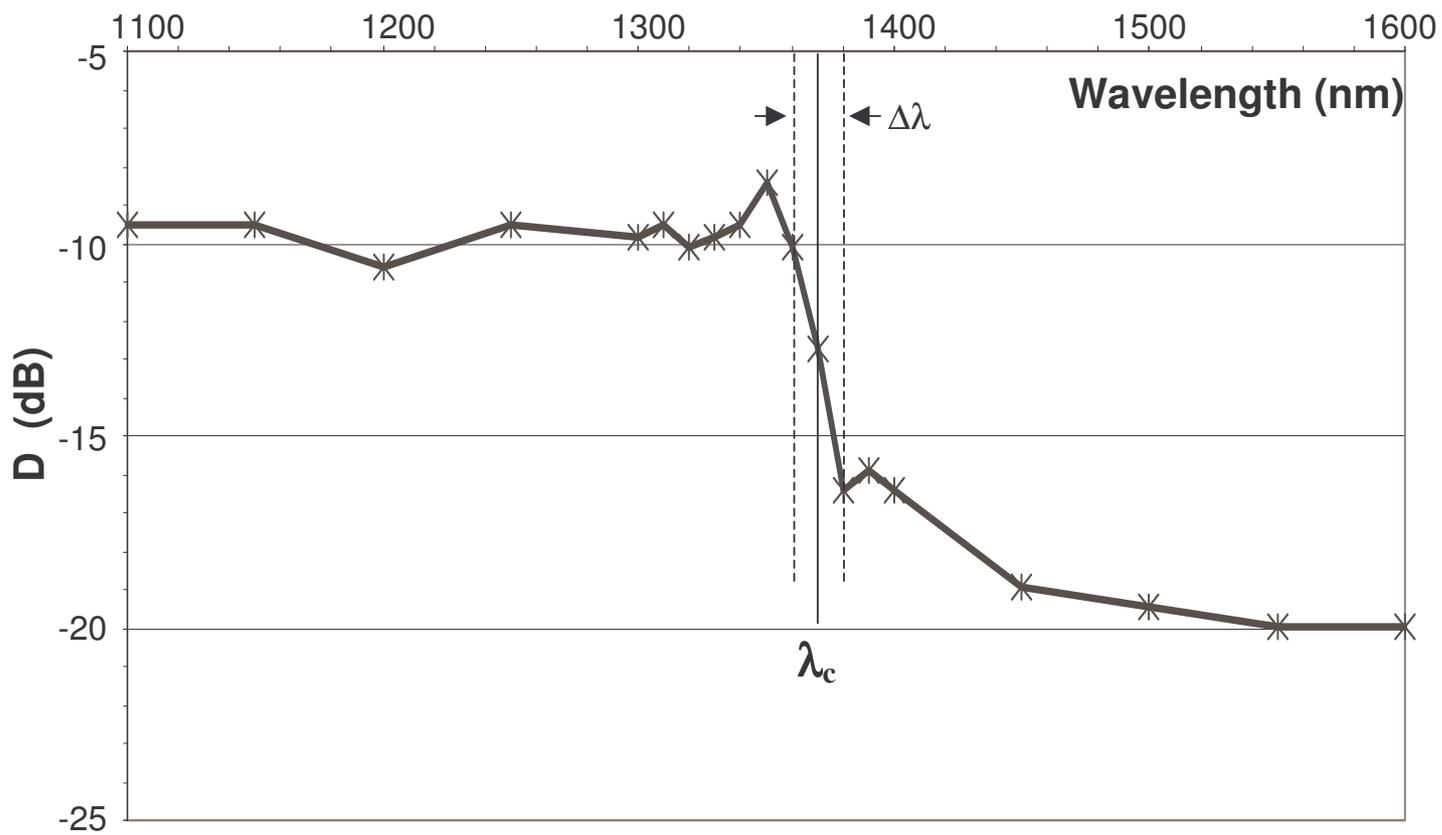

Figure 3

**Figure 4 :** :Q as a function of wavelength for the first higher order modes of the finite size MOF described in the text. For clarity, the curves of the $HE_{21}$ like modes, close to those of the $TE_{01}$ like mode are not presented. The results from the FEM and the MM are shown.

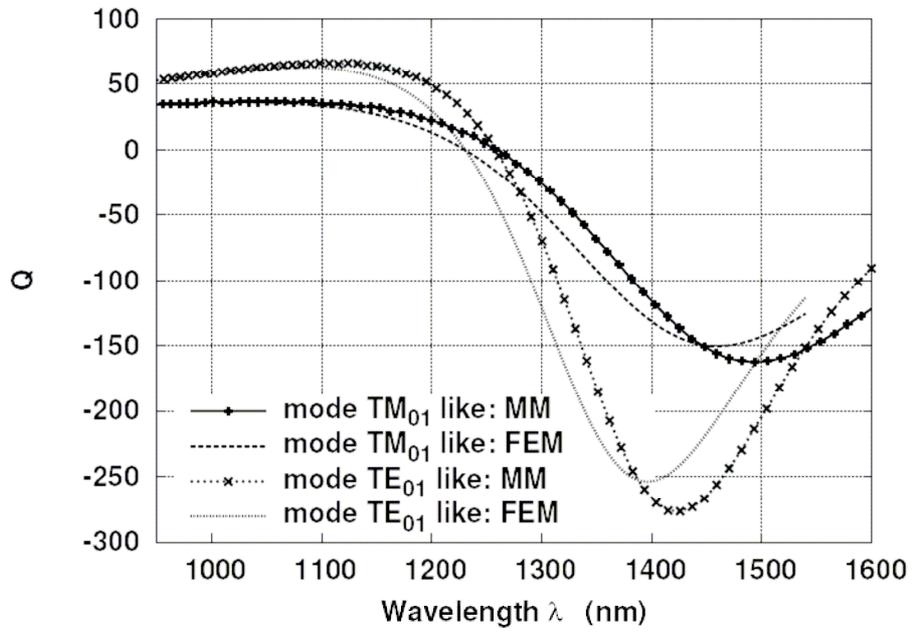

# Figure 4

**Figures captions**

**Figure 1** experimental setup (cross section of the microstructured fiber shown in the inset)

**Figure 2 :** typical spectra measured with the setup described in fig.1 a) single mode regime b) bimode regime ($\Delta=A_c(dBV)-A_l(dBV)$)

**Figure 3 :** Curve $D(\lambda)$ measured with the fiber shown in the fig.1 and described in the text ($\lambda_c$=1370nm, $\Delta\lambda$=20nm)

**Figure 4 :** :Q as a function of wavelength for the first higher order modes of the finite size MOF described in the text. For clarity, the curves of the $HE_{21}$ like modes, close to those of the $TE_{01}$like mode are not presented. The results from the FEM and the MM are shown.